\documentclass[seceqn,10pt]{article}

\usepackage{seceqn}
\usepackage{epsf}
\usepackage{cite}
\usepackage{amsmath}
\usepackage{amsfonts}
\usepackage{amssymb}
\usepackage{bm}
\usepackage{bbm}
\usepackage{graphicx}
\usepackage{epsfig}
\usepackage{latexsym}
\usepackage{nicefrac}
\usepackage{dcolumn}
\usepackage{color}

\newcolumntype{.}{D{x}{}{7}}

\usepackage{fancyhdr}
\def\corresponds{{\lower.1.377ex\hbox{=}}{\rm\kern-.75em^\triangle}}
\def\succsim{\succ\kern-.9em_\sim\kern.3em}
\def\precsim{\prec\kern-1em_\sim\kern.3em}
\def\slantfrac#1#2{\kern1em^{#1}\kern-.3em/\kern-.1em_{#2}}
\def\lfrac#1#2{{}^{#1\!}\kern-.0em/_{#2}}

\def\buildrel#1\under#2{\mathrel{\mathop{\kern0pt #2}\limits_{#1}}}

\definecolor{light}{gray}{0.90}
\definecolor{darker}{gray}{0.50}
\definecolor{dark}{gray}{0.30}

\def\dd{{\mathrm{d}}}
\def\ii{{\mathrm{i}}}
\def\ee{{\mathrm{e}}}

\def\tfrac#1#2{ {\textstyle{\frac{#1}{#2}} } }

\begin{document}

\pagestyle{empty}

\newpage

\vspace*{-2.0cm}
\begin{center}
\begin{tabular}{c}
\hline
\rule[-3mm]{0mm}{12mm}
{\large \sf Lamb Shift in Muonic Hydrogen. ---II.}\\
\rule[-5mm]{0mm}{12mm}
{\large \sf Analysis of the Discrepancy of Theory and Experiment}\\
\hline
\end{tabular}
\end{center}
\vspace{0.0cm}
\begin{center}
U. D. Jentschura\\
\scriptsize
{\em Department of Physics,
Missouri University of Science and Technology,
Rolla, Missouri, MO65409, USA, and\\
National Institute of Standards and Technology,
Gaithersburg, Maryland, MD20899, USA}
\end{center}
\begin{center}
\begin{minipage}{15.0cm}
{\underline{Abstract}}
Currently, both the $g$ factor measurement of the muon as well as the Lamb shift
$2S$--$2P$ measurement in muonic hydrogen are in disagreement with theory.
Here, we investigate possible theoretical explanations,
including proton structure effects and small modifications of the 
vacuum polarization potential.
In particular, we investigate a conceivable small modification of the spectral
function of vacuum polarization in between the electron and 
muon energy scales due to a virtual millicharged particle
and due to an unstable vector boson originating from 
a hidden sector of an extended standard model.
We find that a virtual millicharged particle which 
could explain the muonic Lamb shift
discrepancy alters theoretical predictions for 
the muon anomalous magnetic moment by many standard deviations
and therefore is in conflict with 
experiment. Also, we find no 
parameterizations of an unstable virtual vector boson which
could simultaneously explain both ``muonic'' discrepancies without 
significantly altering theoretical predictions for electronic 
hydrogen, where theory and experiment currently are in excellent agreement.
A process-dependent correction involving electron screening 
% (possibly, in $p \mu^- e^-$ bound systems) 
is evaluated to have the right sign and order-of-magnitude to explain the 
observed effect in muonic hydrogen. 
Additional experimental evidence from light muonic atoms and 
ions is needed in order to reach further clarification.
\end{minipage}
\end{center}

\noindent

Relativistic and quantum electrodynamic (QED) effects in atoms, molecules, and ions;\\
Quantum electrodynamics;\\
Specific calculations;\\
\vfill
\tableofcontents
\vfill
\begin{center}
\begin{minipage}{15cm}
\begin{center}
\hrule
{\bf \scriptsize
\noindent electronic mail: ulj@mst.edu}
\end{center}
\end{minipage}
\end{center}

\newpage

%\clearpage\fancyhead[R]{\normalsize \rightmark}
\pagestyle{fancy}
\sloppy

%
% Introduction
%
\section{Introduction}

Recently, two experiments involving quantum electrodynamics 
(QED) effects have been in disagreement with 
theory. The muon anomalous magnetic 
moment $a_\mu = ( g_\mu - 2 )/2$ has been 
measured~\cite{BeEtAl2002posmuon,BeEtAl2004negmuon,BeEtAl2006muon} as
\begin{equation}
a_{\mu\;{\rm exp}} = 11659208.0(6.3)  \times 10^{-10}
\end{equation}
in $3.4 \, \sigma$ disagreement with some of the latest theoretical
analyses~\cite{HaMaNoTe2007} 
\begin{equation}
a_{\mu\;{\rm th}} = 11659180.4(5.1)  \times 10^{-10} \,.
\end{equation}
The original aim of the recent muonic hydrogen Lamb shift experiment~\cite{PoEtAl2010}
was the determination of the proton radius. When QED theory is assumed 
to be correct, then the value 
\begin{equation}
\label{rpMUON}
r_p = \sqrt{ \left< r^2 \right>_p } = 0.84184(67) \, {\rm fm} 
\end{equation}
is inferred for the root-mean-square proton charge radius 
from a comparison of theory and experiment for the transition 
$2S_{1/2}(F\!=\!1) \Leftrightarrow 2P_{3/2}(F \! =\! 2)$
in muonic hydrogen ($\mu$H).
This value of the proton radius is in disagreement 
with the value obtained in the same way mainly 
from hydrogen and deuterium spectroscopy~\cite{JeKoLBMoTa2005},
which is the basis of the CODATA value~\cite{MoTaNe2008},
\begin{equation}
\label{rpCODATA}
r_p = 0.8768(69) \, {\rm fm}  \,.
\end{equation}
The most recent and accurate measurement of the proton radius 
from electron scattering~\cite{BeEtAl2010}, yields a value of
\begin{equation}
\label{rpSCAT}
r_p = 0.879(8) \, {\rm fm}  \,,
\end{equation}
when the statistical and systematic uncertainties given in 
Ref.~\cite{BeEtAl2010} are added quadratically.
The two values~\eqref{rpCODATA} and~\eqref{rpSCAT} are 
in excellent mutual agreement but differ from the muonic 
hydrogen value~\eqref{rpMUON} by $5.0$~standard deviations.
Consequently, it may be permissible to invert the argument, and to evaluate
current QED theory for the muonic hydrogen transition 
(as summarized in the supplementary material 
published with Ref.~\cite{PoEtAl2010}) with the CODATA 
value~\eqref{rpCODATA}. Using the theoretical expression 
given in Ref.~\cite{PoEtAl2010},
\begin{equation}
\label{Eth_old}
\overline E_{\rm{th}} = 
\left( 209.9779(49) - 5.2262 \frac{r_p^2}{\mathrm{fm}^2}  + 
0.0347 \frac{r_p^3}{\mathrm{fm}^3} \right) {\rm meV} ,
\end{equation}
one obtains, using the CODATA proton radius given in Eq.~\eqref{rpCODATA},
a theoretical prediction of
\begin{equation}
\label{Eth_old1}
\overline E_{\rm{th}} = 205.984(63) \, {\rm fm} \,,
\end{equation}
which is $5.0 \, \sigma$ away from the experimental value of 
\begin{equation}
\label{Eexp}
E_{\rm{exp}} = 206.2949(32) \, {\rm fm} \,,
\end{equation}
reported in Ref.~\cite{BeEtAl2010}.
The recent theory update~\cite{Je2010paper1} 
shifts theoretical predictions only minimally, to
\begin{equation}
\label{Eth_new}
E_{\rm{th}} = 
\left( 209.9974(48) - 5.2262 \frac{r_p^2}{\mathrm{fm}^2} \right) \, {\rm meV} \,.
\end{equation}
Using the CODATA proton radius given in Eq.~\eqref{rpCODATA},
one then obtains the theoretical prediction of
\begin{equation}
\label{Eth_new1}
E_{\rm{th}} = 205.980(63) \, {\rm fm} \,,
\end{equation}
in excellent agreement with~\eqref{Eth_old1},
but in significant disagreement with 
the experimental result given in Eq.~\eqref{Eexp}.

Both of the most recent QED experiments involving
muons~\cite{BeEtAl2006muon,PoEtAl2010} are in disagreement with theory.
The discrepancies have the ``same sign'' and read
\begin{subequations}
\begin{align}
\label{deltaamu}
\delta a_\mu = & \; a_{\mu\;{\rm exp}} - a_{\mu\;{\rm th}} = 
2.76(81) \times 10^{-10} \,,
\\[2ex]
\label{deltaE}
\delta E =& \;
E_{\rm{exp}} - E_{\rm{th}} 
= 0.316(63) \, {\rm{meV}} \,.
\end{align}
\end{subequations}

We here proceed as follows. First, in Sec.~\ref{review},
we present a historical perspective on discrepancies 
in muonic bound systems observed in the past,
and on their eventual resolution. In view of the 
current discrepancy, such a historical perspective 
may be useful.

Possible theoretical 
explanations for the current discrepancy can mainly 
be divided into two categories: proton structure 
effects and modifications of the vacuum polarization
charge density. Hypothetical proton structure effects
are discussed in Sec.~\ref{sec3}, 
and modifications of the vacuum polarization charge 
density in Sec.~\ref{sec4}. 
Section~\ref{sec3} is divided into two subsections,
the first of which deals with a conceivable ``dip''
in the proton form factor slope in the momentum 
transfer range studied in muonic hydrogen spectroscopy,
and the second deals with a conceivable, anomalously large
contribution from the inelastic part of the two-photon
exchange diagram. 
Section~\ref{sec4} is divided into three parts.
These deal with a conceivable nonperturbative 
correction to the vacuum polarization
potential, with the contribution of a 
millicharged particle that modifies the 
vacuum polarization loop, and with the contribution 
of a conceivable virtual, unstable vector boson that modifies the 
vacuum polarization potential.
We anticipate here that none of these considerations
will lead to a definitive candidate for an 
explanation of the discrepancy.
Still, a compilation of a number of possible
explanations appears to be useful in the current 
situation. Some of the discussed explanations 
are relevant only to muonic hydrogen, which is 
the main subject of the current article,
others may be relevant for both observed discrepancies.
Possible electron screening corrections are discussed in Sec.~\ref{straw}.
Conclusions are reserved for Sec.~\ref{conclu}.
Natural units with $\hbar = c = \epsilon_0 = 1$ are used throughout 
the paper. 

%
% Perspectives on the Disagreement in Muonic Hydrogen
%
\section{Historical Perspective on Discrepancies
in Muonic Systems}
\label{review}

Let us briefly comment on the size of the 
disagreement in muonic hydrogen 
[see Eqs.~\eqref{Eth_new1} and~\eqref{Eexp}].
Current predictions are based on the 
calculations reported in 
Refs.~\cite{Pa1995,Pa1999,Pa2007,Bo2005mup,Bo2005mud,FaMa2003,Ma2005mup,Ma2007mup}
and represent the result of independent groups. 
Important contributions originally 
calculated in Refs.~\cite{Pa1995,Pa1999} 
have been verified in Refs.~\cite{Bo2005mup,Bo2005mud}.
Higher-order vacuum polarization effects have been 
given special attention in Refs.~\cite{Bo2005mup,Bo2005mud}.
The theory used in the evaluation of the experiment~\cite{PoEtAl2010} has been compiled 
at Laboratoire Kastler--Brossel in Paris.
The disagreement of theory and experiment is on the level of 
$\tfrac15$ of the two-loop vacuum polarization 
correction which amounts to $1.508 \, {\rm meV}$ and which 
was calculated first by Kallen and Sabry~\cite{KaSa1955},
then recalculated in Ref.~\cite{BaRe1973}.
A clear exposition is given 
in volume III of Ref.~\cite{Sc1970}
(the result has later been generalized to non-Abelian 
gauge theories, see Refs.~\cite{Kn1990,BrFl1993,ChKuSt1996}).

One may point out that the discrepancy~\eqref{deltaE}
amounts to (roughly) 1.5~parts per thousand
of the total vacuum polarization effect in muonic hydrogen.
By contrast, in a previous experiment~\cite{BeEtAl1986} 
involving muonic $3d \Leftrightarrow 2p$
transitions in ${}^{24}$Mg and ${}^{28}$Si,
the vacuum polarization effect has already been verified
to 1.0~parts per thousand
(a relative accuracy of $950\times 10^{-6}$ is quoted in Ref.~\cite{BeEtAl1986})
and thus, to better precision than the current disagreement.
If there were any fundamental reason 
for a deviation of theory and experiment on this 
level, then one might wonder if the effect 
(whichever it is) might have been visible in 
the experiment reported in Ref.~\cite{BeEtAl1986}. 
However, the muonic transition reported in Ref.~\cite{BeEtAl1986}
suffers from an uncertainty due to electron screening, 
and also, as it involves non-$S$ states, the overlap of the 
muonic wave functions with the nucleus are not as pronounced
as for $S$ states.
So, the quoted experiment~\cite{BeEtAl1986} is 
not sensitive to higher-order nuclear structure effects,
and it also probes the vacuum polarization at a different energy scale
as compared to $S$ states which are much closer to the nucleus.

In the 1970s, experiments involving muonic transitions
were found to be in disagreement with theory~\cite{DiEtAl1971}.
Part of the discrepancies were addressed after
a sign error in the calculation of the two-loop
vacuum polarization correction~\cite{Fr1969} was
eliminated~\cite{Bl1972}. An elucidating discussion of the 
status reached in 1978 is given in Ref.~\cite{BrMo1978}.
Further clarification was reached
when a standard $\gamma$-ray spectrometer used in the experiments
was recalibrated~\cite{DeKeSaHe1980}.
A few remarkable experiments later found full agreement 
of theory and experiment in muonic systems
(e.g., Refs.~\cite{TaEtAl1978,BeEtAl1986}).
In Ref.~\cite{DuEtAl1974}, the nuclear radii of 
some carbon, nitrogen and oxygen isotopes are determined 
by analyzing muonic transitions, and the resulting radii
are in agreement with electron scattering radii for the 
investigated nuclei to better than $5\,\%$.
Later, the radius of ${}^{12}{\rm C}$ has been 
updated~in Refs.~\cite{RuEtAl1984,OfEtAl1991},
converging at a value of
$r_C = 2.478(9) \, {\rm fm}$,
in excellent agreement with the value from
muonic x-ray studies.
After this finding, a conceivable non-universal coupling of the 
electron versus muon to the nucleus was discarded,
and muonic spectroscopy meanwhile is an established tool
for the determination of nuclear radii~\cite{AnEtAl2009}.

Indeed, the current disagreement of the proton radius 
derived from muonic versus  electronic hydrogen radius amounts 
to about $4\,\%$ and is a large discrepancy
(roughly, electron scattering and hydrogen spectroscopy gives a result
of $r_p \approx 0.88\,{\rm fm}$, whereas the recent 
muonic Lamb shift experiment yields $r_p \approx 0.84\,{\rm fm}$).
Therefore, it is permissible to speculate about nonperturbative 
effects and new physics effects for both discrepancies
(muon $g$ factor and muonic hydrogen Lamb shift).
Among the two effects, the hydrogen experiment is perhaps
the most interesting, (i) because its theory, on the level
of the discrepancy, is given by only few, simple bound-state QED effects
not exceeding the two-loop level
(see Ref.~\cite{Je2010paper1}), and (ii) because new experiments
in related systems are planned which may or may not confirm the 
observed discrepancy~\cite{PoPriv2010}. Here, we thus 
investigate a few of these possible explanations.

%
% Proton Structure Effects
%
\section{Proton Structure Effects}
\label{sec3}

%
% Form Factor
%
\subsection{Form Factor}
\label{sec31}

The proton mean-square charge radius is defined
in terms of the slope of the Sachs $G_E$ form factor 
of the proton,
\begin{equation}
\langle r^2 \rangle = 6 \, 
\left. \frac{\partial G_E(q^2)}{\partial q^2} \right|_{q^2 = 0}
= -6 \, 
\left. \frac{\partial G_E(Q^2)}{\partial Q^2} \right|_{Q^2 = 0} \,,
\end{equation}
where $Q^2 = -q^2$ is the space-like momentum transfer.
Different ranges of the momentum transfer
are relevant for the calculation of the slope 
in different experiments.
The proton radius from electronic hydrogen is 
determined from exchanged Coulomb photons with
momentum transfers in the region
\begin{equation}
Q^2 \sim (\alpha m_e c)^2 = \left( 3.7 \times 10^{-6} \, \frac{\rm GeV}{c} \right)^2 \,.
\end{equation}
For muonic hydrogen, the atomic momentum is in the range of
\begin{equation}
\label{probemuon}
Q^2 \sim (\alpha m_\mu c)^2 = \left( 7.7 \times 10^{-4} \, \frac{\rm GeV}{c} \right)^2 \,.
\end{equation}
One may point out that 
this is just below the electron-positron pair production threshold,
\begin{equation}
\label{etransfer}
(2 m_e c)^2 = \left( 1.0 \times 10^{-3} \, \frac{\rm GeV}{c} \right)^2 \,.
\end{equation}
The momentum transfer range probed in the 
recent electron scattering experiment~\cite{BeEtAl2010} is 
larger, but not excessively larger,
\begin{equation}
\left( 6.3 \times 10^{-2} \, \frac{\rm GeV}{c} \right)^2 < Q^2
 < \left( 1 \, \frac{\rm GeV}{c} \right)^2 \,.
\end{equation}
The slopes of the proton form factor determined from the 
electron scattering data and from electronic hydrogen 
spectroscopy are in excellent mutual agreement.
The momentum transfer range for muonic hydrogen 
spectroscopy lies in between these two ranges. 
Consequently, it would be somewhat surprising if the
proton form factor slope had a ``dip'' in this range 
that would explain the discrepancy for muonic hydrogen.
Still, without a direct scattering measurement in this
momentum transfer range, this possibility cannot
be fully excluded at present.

%
% Proton Polarizability
%
\subsection{Proton Polarizability}

Generically, the proton polarizability can
be related to the resonances of the proton (its excitation spectrum) via
dispersion relations [see Eqs.~(29) and~(30) of Ref.~\cite{Pa1999}],
and related to the inclusive reaction $e + p \to e' + X$.
This is an accepted procedure for all nuclei, also for 
heavier nuclei~\cite{DuEtAl1974}.  For the proton, one would
intuitively assume that the bulk of the contribution is from the
$\Delta(1232)$ resonance, which has been measured well. Yet, the data
obtained in the literature for the proton polarizability contribution
scatter, and in Ref.~\cite{PoEtAl2010},
the contribution is currently estimated
as $+ 0.015(4) \, {\rm meV}$ 
based on the scatter of values obtained from different 
theorists~\cite{Pa1999,Ro1999,MaFa2000}. 
The proton polarizability correction
to the $2P$--$2S$ Lamb shift has been calculated as
$0.0012\,{\rm meV}$ in Ref.~\cite{Pa1999}, which is 
an order to magnitude smaller than the discrepancy $\delta E$.
Other authors~\cite{Ro1999,MaFa2000} confirm the magnitude
of the result and give values of $0.0015\,{\rm meV}$ for the
proton polarizability correction to the Lamb shift in
muonic hydrogen. 
It would thus be helpful to reevaluate the effect,
and to obtain more accurate estimates, even if the 
current uncertainty estimate of $\pm 0.004 \, {\rm meV}$ 
is numerically tiny as compared to the 
discrepancy of $\delta E \approx 0.31 \, {\rm meV}$.

In addition, one may point out that in the past, nuclear
radii inferred from muonic transitions and from
electron scattering have agreed to better than $5\,\%$
(see the discussion in Sec.~\ref{review}).
So, if an anomalously large proton polarizability contribution
were found by a reanalysis, then one might have to revisit
this effect also for other bound systems and in the more
general context of the validity of the nuclear charge radius determination
from muonic transitions~\cite{An2004,AnEtAl2009}.

%
% Vacuum Polarization
%
\section{Vacuum Polarization}
\label{sec4}

%
% Nonperturbative Vacuum Polarization
%
\subsection{Nonperturbative Vacuum Polarization}
\label{sec41}

Let us briefly review why a nonperturbative vacuum 
polarization effect might have been considered as 
an explanation of the discrepancy in muonic 
hydrogen. The spectrum of muonic hydrogen is influenced 
by electronic vacuum polarization effects.
The muon is heavier than the 
electron by a factor of $m_\mu/m_e \approx 207$,
and the reduced mass of muonic hydrogen is roughly equal
to the muon mass. The effective Bohr radius in muonic hydrogen 
is $1/(\alpha \, m_R) = 284.748 \,{\rm fm}$,
which is smaller than the reduced Compton wavelength of the 
electron, $1/m_e = 386.159 \, {\rm fm}$.
The bound muon thus enters the 
electronic vacuum polarization charge cloud of the proton.
The electronic vacuum polarization shift in muonic hydrogen
is of order $\alpha^3 m_R$, where $m_R$ is the 
reduced mass,  and thus more pronounced than 
the electronic vacuum polarization shift in electronic 
hydrogen, where the vacuum polarization contribution
to the Lamb shift is of order $\alpha^5 m_R$.
The large vacuum polarization shift is also responsible
for the fact that the $2S$ level in muonic 
hydrogen is energetically lower than the $2P_{1/2}$ state
(in contrast to electronic hydrogen, where the situation 
is opposite).

Superficially, the vacuum polarization effects
converge very well in terms of the QED loop expansion.
The one-loop effect gives a
contribution of 
$205.0074 \, {\rm meV}$ to the $2P$--$2S$ Lamb shift,
in first-order perturbation theory, 
while the second-order effect adds
$0.1509 \, {\rm meV}$.
The two-loop (Kallen--Sabry) shift is $1.5081 \, {\rm meV}$, 
followed by the higher-order Wichmann-Kroll term of 
$-0.00103 \, {\rm meV}$.
Eventually, of course, the expansion will diverge
according to a famous argument put forward by Dyson~\cite{Dy1952},
but the expected nonperturbative effect would be
of order $\exp(-1/\alpha)$ and thus completely 
negligible for the muonic hydrogen experiment.

However, the superficial convergence still does not 
exclude the presence of a much larger nonperturbative 
correction to the vacuum polarization {\em if} the 
{\em local} convergence of the loop expansion of the 
higher-order vacuum polarization potentials
breaks down in close vicinity of the proton,
i.e., if the higher-order (Kallen--Sabry and Wichmann--Kroll) terms 
are more singular than the one-loop Uehling term for $r \to 0$.
In that case, a nonperturbative correction 
to the vacuum polarization potential might have led
to a highly nonlinear, nonperturbative correction and 
one might have had to solve the Schr\"{o}dinger equation using the 
full nonperturbative potential near the origin.
In that case, lattice methods would probably have had to be 
invoked in order to calculate the full vacuum 
polarization potential for $r \to 0$.

The question whether this more elaborate calculation is 
necessary, can only be answered by a concrete 
calculation of the leading asymptotics of the 
Uehling, Kallen--Sabry, and Wichmann--Kroll potentials 
for $r \to 0$. One finds, in agreement with Ref.~\cite{Bl1972},
for the leading asymptotics of the one-loop Uehling 
potential $V_{\rm vp}(r)$,
\begin{equation}
V_{\rm vp}(r) \sim \frac{2 \,\alpha^2}{3 \,\pi r} \, \ln(m_e \,r) \,,
\qquad r \to 0\,,
\end{equation}
for the Kallen--Sabry potential,
\begin{equation}
V_{\rm KS}(r) \sim -\frac{4 \, \alpha^3}{9 \, \pi r} \, \ln^2(m_e \,r) \,,
\qquad r \to 0\,,
\end{equation}
and for the Wichmann--Kroll potential,
\begin{equation}
V_{\rm WK}(r) \sim \frac{\alpha^4}{\pi r} \, 
\left( - \frac23 \, \zeta(3) + \frac16 \pi^2 - \frac79 \right) \,,
\quad r \to 0\,.
\end{equation}
These potentials have to be compared to the Coulomb
potential
\begin{equation}
V(r) = -\frac{\alpha}{r} \,.
\end{equation}
By inspection of these formulas,
we conclude that the higher-order vacuum-polarization potentials
are of the same order-of-magnitude  as the Coulomb potential for 
distances shorter than
\begin{equation}
r \sim \frac{\exp(-1/\alpha)}{m_e} = 1.2 \times 10^{-57} \, {\rm fm}\,,
\end{equation}
which is the length scale of the Landau pole.
This length scale is not sufficient to induce to any conceivably 
large nonperturbative effects.

\begin{figure}[t!]
\begin{center}
\begin{minipage}{0.7\linewidth}
\begin{center}
\includegraphics[width=0.91\linewidth]{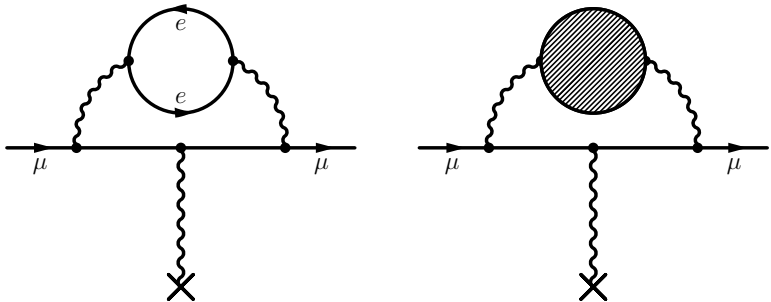}
\end{center}
\caption{\label{fig2} Vertex correction to the muon 
anomalous magnetic moment with an additional
vacuum-polarization insertion in the photon line
(left: electron-positron loop, right: non-QED virtual 
process).}
\end{minipage}
\end{center}
\end{figure}

\begin{figure}[t!]
\begin{center}
\begin{minipage}{0.7\linewidth}
\begin{center}
\includegraphics[width=0.6\linewidth]{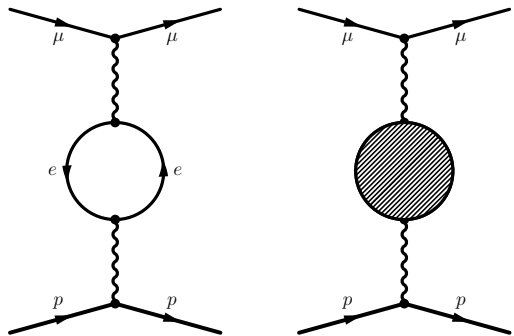}
\end{center}
\caption{\label{fig3} Vacuum polarization correction to the 
muonic hydrogen Lamb shift 
(left: electron-positron loop, right: the blob denotes a
correction due to a non-QED virtual
process, e.g., hadronic vacuum polarization, $\rho(770)$ meson pole,
or due to a  hypothetical low-energy vector meson).}
\end{minipage}
\end{center}
\end{figure}

%
% Virtual Millicharged Particles
%
\subsection{Virtual Millicharged Particles}
\label{sec42}

One of the most interesting possibilities for an explanation of both ``muonic
QED discrepancies'' observed at present would be due to the contribution of a
virtual millicharged particle.  A millicharged particle was invoked as a
possible explanation for the observed (later retracted) 
optical rotation~\cite{ZaEtAl2006,ZaEtAl2008paper1,BrEtAl2008paper2} of
linearly polarized laser light by a magnetic field.  If the photon initiates
pair production of light charged fermions with masses below the electron mass
and charge on the order of $q = \epsilon \,e$ with $\epsilon \ll 1$, then the
initial observation made in Ref.~\cite{ZaEtAl2006} could be explained
(see Refs.~\cite{GiJaRi2006epl,GiJaRi2006prl,AhGiJaRi2007}).
The non-integer charge does not contradict
charge quantization if the millicharged particles are generated from a
``hidden'' sector of the standard model via the Stueckelberg
mechanism~\cite{ChYu2001,FeLiNa2007}.  Such millicharged particles have been
searched in devoted experiments at SLAC~\cite{PrEtAl1998}.  Some of these
experiments are sensitive only to {\em stable} millicharged particles, because
they depend on obtaining a signal from particle detectors, as pointed out in
Ref.~\cite{La1998}.  Several bounds have been derived regarding the mass and
charge of such particles, which have otherwise been quoted as a candidate for
dark matter (see
Refs.~\cite{DaCaBa1991,DuGo2001,DaHaRa2006,MiRaSe2005,MiRaSe2007,HoSe2007}).

The muon anomalous magnetic moment discrepancy and the Lamb shift discrepancy
have the same sign, i.e., the experimental result is larger than the
theoretical prediction.  An additional virtual excitation of a quantum field (a
virtual particle) would naturally be assumed to enhance both effects.  The muon
anomalous magnetic moment is numerically small, the correction induced by a
hypothetical virtual particle is a two-loop effect (see Fig.~\ref{fig2}),
whereas for the muonic hydrogen Lamb shift, the conceivable contribution of a
millicharged particle only is a one-loop correction (see Fig.~\ref{fig3}).
Therefore, it is indicated to map out possible parameter ranges for the
hypothetical millicharged particle.  In muonic experiments, one is very
sensitive to the mass range $m_e \ll m_M \ll m_\mu$ for hypothetical virtual
particles. If the virtual particle is in this mass range, then the effect on
the muon anomalous magnetic moment, and on the muonic Lamb shift is  enhanced
because $m_M \ll m_\mu$, but suppressed for electronic systems such as ordinary
hydrogen because $m_e \ll m_M $. These models are simple-minded,
straightforward {\em ansatz}es that ``suggest 
themselves'' because of the pertinent mass region.
We thus restrict the discussion to conceivable millicharged 
particles and do not consider supersymmetric graphs in which, e.g, the 
muon might turn into a virtual smuino, emitting a charged higgsino or wino.
We also do not consider hypothetical corrections from
axion electrodynamics~\cite{Wi1987axion}.

Furthermore, since $m_e \ll m_M$ by
assumption, the modification of the vacuum polarization due to the millicharged
particle, for electronic hydrogen, can be absorbed into a Dirac $\delta$
potential acting on the electronic hydrogen wave functions.  Its functional
form is therefore indistinguishable from the nuclear finite size effect for
electronic hydrogen and could be ``absorbed into'' a modification of the proton
radius inferred from electronic hydrogen spectroscopy without any further
observable consequences for atomic transitions in electronic hydrogen. For
muonic hydrogen, however, since $m_M \ll m_\mu$, the hypothetical millicharged
particle leads to an enhanced energy correction, different from a Dirac
$\delta$.  The mass range $m_e \ll m_M \ll m_\mu$ therefore is the primary
parameter range probed by muonic QED experiments.
For $m_M \gg m_\mu \gg m_e$, the effect of the millicharged 
particle amounts to a Dirac $\delta$ function and is thus 
indistinguishable from the contribution of the nuclear size effect
for {\em both} muonic as well as electronic hydrogen.

We thus proceed as follows. First, we find a convenient parameterization of the
expected modification of the spectral function of vacuum polarization, as a
function of the charge and mass of the millicharged particle. Then, evaluate
the shift of the anomalous magnetic moment of the muon and of the Lamb shift
due to the millicharged particle (these are both proportional to the square of
the charge), and relate them to the observed discrepancies, as a function of
the mass of the assumed millicharged particle.  Again, forming the ratio of
these relative shifts, we investigate if there is a parameter range for which
both the anomalous magnetic moment of the muon and the discrepancy observed in
muonic hydrogen could be explained by the virtual particle.

% We adopt a naive point of view and ask ourselves:
% what modification of the 

First, let us find a convenient parameterization for the 
spectral density of vacuum polarization. 
As is well known,
the effect of electronic vacuum polarization
on the photon propagator can be described by the replacement
\begin{align}
\frac{1}{q^2 + \ii\epsilon} \to & \;
\frac{\alpha}{3 \pi} \int_{4 m_e^2}^\infty \frac{\dd t}{t} \,
\rho_e(t) \frac{1}{q^2 - t + \ii\epsilon} \,, 
\nonumber\\[2ex]
\rho_e(t) = & \; 
\sqrt{ 1 - \frac{4 m_e^2}{t} } \,
\left( 1 + \frac{2 m_e^2}{t} \right)\,,
\end{align}
in the photon propagator,
where we refer to $\rho_e(t)$ as the spectral density.
The corresponding vacuum polarization potential
(Uehling potential) induced by electronic vacuum polarization is
\begin{equation}
V_{\rm vp}(r) = -\frac{\alpha^2}{3 \pi} \,
\int_{4 m_e^2}^\infty \frac{\dd t}{t} \, 
\frac{\ee^{-\sqrt{t} \, r}}{r} \, \rho_e(t) \,.
\end{equation}
The spectral density is zero at threshold $t = 4 m_e^2$ and quickly
approaches the asymptotic value $\rho_e(t) \to 1$ for $t \to \infty$.
Although the millicharged particle has been assumed to be
of spin $1/2$, we emphasize here that similar
threshold behavior can be expected regardless of the
spin of the millicharged particle~\cite{GlRaRe2007,ItZu1980}.

\begin{figure}[t!]
\begin{center}
\begin{minipage}{0.7\linewidth}
\begin{center}
\includegraphics[width=0.91\linewidth]{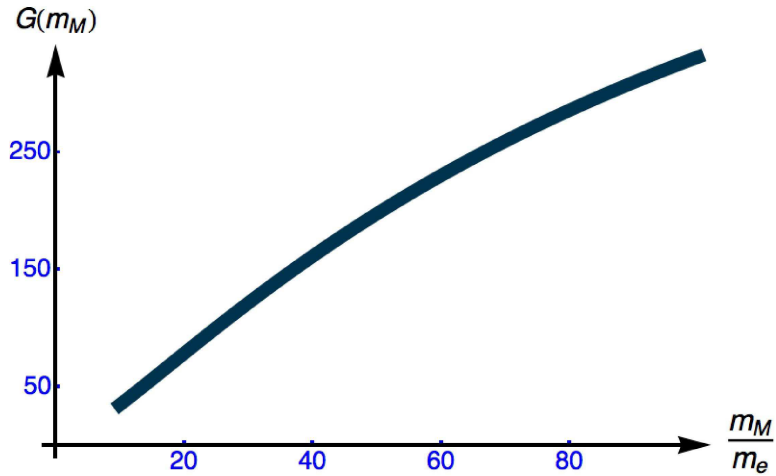}
\end{center}
\caption{\label{abb4} (Color online.) In the range 
$10 \,m_e < m_M < 100 \,m_e$, the function $G(m_M)$ is 
a lot larger than unity, as shown in the plot.
If a hypothetical millicharged particle
in the given mass range were responsible for the 
discrepancy observed in muonic hydrogen, then the 
same particle would lead to complete disagreement for the 
anomalous magnetic moment of the muon.}
\end{minipage}
\end{center}
\end{figure}

\begin{figure}[tt!!]
\begin{center}
\begin{minipage}{0.7\linewidth}
\begin{center}
\includegraphics[width=0.91\linewidth]{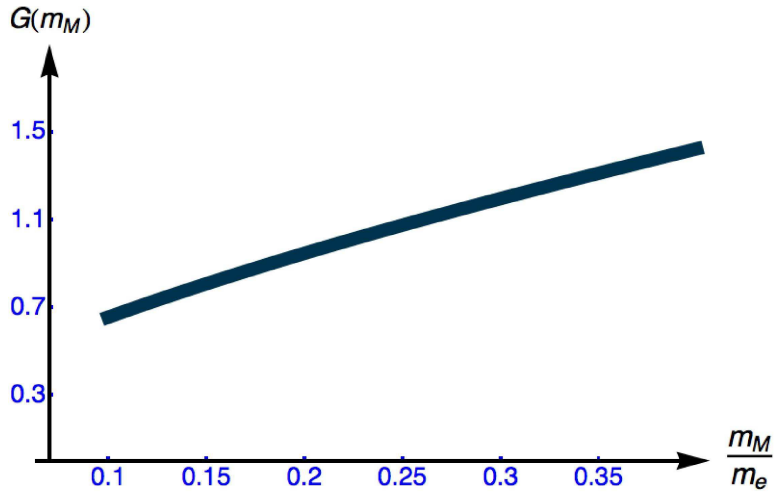}
\end{center}
\caption{\label{abb5} (Color online.) 
Both the muon anomalous magnetic moment discrepancy as well as 
the muonic hydrogen Lamb shift discrepancy can be explained 
by a millicharged particle of mass $m_M = 0.221 \, m_e$ and 
charge $q = \pm 0.0179 e$, as shown in the graph.
Indeed, one finds  $G(0.221 \, m_e) = 1$. However, in the indicated
mass range, the correction to the electronic hydrogen Lamb shift 
induced by the millicharged particle becomes so large that it 
leads to an inconsistent, sizeable shift of the the proton
radius inferred from the hydrogen Lamb shift. See text for
further explanations.}
\end{minipage}
\end{center}
\end{figure}

A millicharged particle modifies the 
spectral density of vacuum polarization 
according to $\rho_e(t) \to \rho_e(t) + \delta\rho(t)$.
For a spin-$1/2$ millicharged particle,
of charge $q = \epsilon \, e$ and mass $m_M$,
we can approximate this modification as
\begin{equation}
\label{deltarho}
\delta \rho(t) \approx \epsilon^2 \, \Theta(t - 4 m_M^2)
\end{equation}
where $\Theta$ is the step function.
We intend to compare changes in the anomalous magnetic 
moment of the muon and in the Lamb shift induced by the 
millicharged particle.
In line with intuition, we here need a positive
spectral function $\delta \rho(t)$, increasing the Lamb shift
for muonic systems and increasing the
muon $g$ factor. This is the right sign because
the theoretical prediction for the
muon $g$ factor as well as the theoretical prediction
for the muonic helium Lamb shift are lower than
the corresponding experimental results.

The correction to the 
anomalous magnetic moment due to electronic 
vacuum polarization is~\cite{LaPedR1972,Ad1974prd1}
\begin{equation}
\delta a_\mu = 
\frac{\alpha^2}{6 \pi^2} \, 
\int_{4 m_e^2}^\infty \frac{\dd t}{t} \, 
f_a(t) \, \rho_e(t) \,,
\end{equation}
where for the muon
\begin{equation}
f_a(t) = 2 \int_0^1 \dd x \, 
\frac{x^2 \, (1-x)}{x^2 + (1-x) \, t/m_\mu^2} \,.
\end{equation}
We have checked that
if one replaces in this expression $m_\mu \to m_e$ 
and integrates over $t$, then one obtains the 
known contribution~\cite{BaMiRe1972},
\begin{equation}
\delta a_e = \frac{139}{36} - \frac{\pi^2}{3} \,,
\end{equation}
to the electron anomalous magnetic moment $a_e$,
due to the diagram on the left in Fig.~\ref{fig2}.

Although the integral representation of $f_a(t)$ is compact,
the analytic result requires us to differentiate two cases,
depending on whether $t < 4 m_\mu^2$ or $t > 4 m_\mu^2$.
For $0 < t < 4 m_\mu^2$, with $\tau = t/(4 m_\mu^2)$,
one finds
\begin{subequations}
\begin{align}
f_a(t) =& \; 1 - 8 \, \tau - 8 \, \tau ( 1 - 2\tau)\, \ln(4 \tau) 
- 4 (1 - 8\, \tau + 8\, \tau^2) \, 
\left( \frac{\tau}{1 - \tau} \right)^{1/2} \,
\arctan\left( \sqrt{\frac{1- \tau}{\tau} } \right) \,,
\end{align}
whereas above the muon threshold, for $t \geq 4 m_\mu^2$, with
$x = \left(1 - \sqrt{1 - 4 m_\mu^2/t}\right)
\bigg/\left(1 + \sqrt{1 - 4 m_\mu^2/t}\right)$,
the result for $f_a(t)$ is
\begin{align}
f_a(t) =& \; x^2 \, (2 - x^2) +
2 \, (1 + x)^2 \, (1 + x^2 ) \, 
\frac{ \ln(1 + x) - x + \tfrac12 \, x^2 }{ x^2 } 
+ 2 \, x^2 \, \frac{1 + x}{1 - x} \, \ln(x) \,.
\nonumber
\end{align}
\end{subequations}
Consequently, the contribution to the muon anomalous 
magnetic moment due to the hypothetical millicharged particle,
divided by the observed discrepancy $\delta a$ given in 
Eq.~\eqref{deltaamu}, is
\begin{equation}
\chi_a = \int_{0}^\infty \frac{\dd t}{t} \, 
\eta_a(t) \, \delta \rho(t) \,,
\end{equation}
where
\begin{equation}
\eta_a(t) = \frac{\alpha^2}{6 \pi^2} \, \frac{f_a(t)}{ \delta a} \,.
\end{equation}
Here, $\delta \rho(t)$ is the vacuum polarization spectral
density given in Eq.~\eqref{deltarho} due to the 
millicharged particle.

For the muonic hydrogen Lamb shift, the situation is as follows.
The one-loop electronic vacuum polarization shifts the $2S$ level downward
by $-219.6\,{\rm meV}$, whereas the $2P$ is shifted downward by 
only $-14.6\,{\rm meV}$. This is because of the enhanced 
probability density of the $2S$ state as compared to $2P$, 
near the nucleus. The total effect on the 
Lamb shift, by both electronic vacuum polarization and also 
by a hypothetical millicharged particle, can thus be approximated 
by taking the negative of the vacuum polarization 
energy shift of the (energetically lower) $2S$ level.
When the resultant shift is divided by 
the observed discrepancy $\delta E$ given in Eq.~\eqref{deltaE},
one obtains the ratio
\begin{equation}
\chi_\mu = 
\int_{0}^\infty \frac{\dd t}{t} \, \eta_\mu(t) \, \delta \rho(t) \,.
\end{equation}
where, again with $\tau = t/(4 m_\mu^2)$,
\begin{align}
\eta_\mu(t) = & \;
\frac{\alpha}{3 \pi} \, \frac{1}{\delta E}\,
\left< 2S \left| 
\frac{\alpha}{r} \; \ee^{-\sqrt{t} \, r}
\right| 2S \right>
= \frac{\alpha^3 \, m_R}{12 \pi} \, \frac{1}{\delta E}\,
\left( 1 + \frac{8 v^2 \tau}{\alpha^2} \right) \,
\left( 1 + \frac{2 v \sqrt{\tau}}{\alpha} \right)^{-4} \,.
\end{align}
Here, $v = m_e/m_R$ is ratio of the electron mass to the 
reduced mass of the muonic hydrogen system.
The calculation of the ratio
\begin{equation}
G(m_M) = 
\frac{\chi_a}{\chi_\mu} 
= 
\frac{\displaystyle \int_{0}^\infty \frac{\dd t}{t} \, \eta_a(t) \, \delta \rho(t)}%
{\displaystyle \int_{0}^\infty \frac{\dd t}{t} \, \eta_\mu(t) \, \delta \rho(t)}
\end{equation}
then answers the following question: Suppose that the 
energy discrepancy $\delta E$ in muonic hydrogen were due to the millicharged 
particle, then how much would the muon anomalous magnetic moment 
be changed by that same millicharged particle, in terms of the 
observed discrepancy $\delta a$? Within the approximation~\eqref{deltarho},
the ratio $G$ depends only on the mass (not on the charge) of the millicharged 
particle,
\begin{equation}
\label{defG}
G(m_M) = 
\frac{\displaystyle \int_{4 m_M^2}^\infty 
\frac{\dd t}{t} \, \eta_a(t) }%
{\displaystyle \int_{4 m_M^2}^\infty 
\frac{\dd t}{t} \, \eta_\mu(t) } \,,
\end{equation}
because $\epsilon^2$ as given in Eq.~\eqref{deltarho} cancels.
If we could find $m_M$ so that $G(m_M) = 1$, then a millicharged
particle would be a serious candidate to explain both 
the muonic anomalous magnetic moment as well as the muonic Lamb shift
discrepancy.

The somewhat disappointing result of a numerical study of $G(m_M)$ is given in
Fig.~\ref{abb4}.  In the mass range $10 \, m_e < m_M < 100\,m_e$, the function
$G(m_M)$ is in the range $25 < G(m_M) < 300$.  The value $G(m_M) = 25$ implies
that, if for given mass of the millicharged particle, the muonic hydrogen Lamb
shift discrepancy is resolved, then we induce a discrepancy of theory and
experiment for the muon anomalous moment by roughly $25 \times 3.4 =
85$~standard deviations.  Expressed differently, the correction to the muon
anomalous magnetic moment, expressed in units of the observed discrepancy
$\delta a_\mu$, is larger by at least a factor $25$ than the modification of
the muonic Lamb shift induced by the millicharged particle, expressed in units
of the observed discrepancy $\delta E$.  That means that a millicharged particle
in the given mass range cannot explain both observed discrepancies.  The
observed discrepancy $\delta E$ is too large to be explained by a millicharged
particle with $m_e \ll m_M \ll m_\mu$ because this would induce a prohibitively
large modification of $\delta a_\mu$.  Conversely, if a 
virtual millicharged particle
provides an explanation for the observed discrepancy $\delta a_\mu$, then it
will explain at most $4 \,\%$ of $\delta E$.  Thus, a millicharged particle in
the given mass range might still explain the discrepancy $\delta a_\mu$, but if
that assumption is true, then the bulk of the explanation for $\delta E$ has to
come from a different effect (e.g., proton structure).

We have previously stressed that muonic QED experiments 
are especially sensitive to a mass range $m_e \ll m_M \ll m_\mu$
of the hypothetical particle. In principle, 
one might still explore the possibility of a millicharged particle with mass
$\alpha m_e \ll m_M \ll m_e$, because in that mass range,
the correction to the hydrogen Lamb shift
is still expressible in terms of a Dirac $\delta$ function and therefore can be
absorbed into a modified proton radius~\cite{MoTaNe2008}.
This study is indicated 
even if the mass range $m_M \ll m_e$ is not the primary range tested by muonic 
experiments. According to Fig.~\ref{abb5},
we find that, in principle, a particle
with $\epsilon \approx 0.0179$ and $m_M \approx 0.221 \, m_e$ could explain both observed
discrepancies $\delta E$ and $\delta a$, simultaneously. 

However, this hypothetical particle is excluded for two reasons.  First, a
calculation of its contribution to the {\em electron} anomaly
would be $\delta a_e = 2.3 \times 10^{-10}$ which is 
much larger than the experimental uncertainty of $\pm 2.8 \times 10^{-13}$
of the recent measurement~\cite{HaFoGa2008}.
A shift in the electron anomaly by
$\delta a_e = 2.3 \times 10^{-10}$ would lead to a relative shift of the
fine-structure constant by $2.0 \times 10^{-7}$ and thus, to 
a severe disagreement with
other determinations of $\alpha$ (see Ref.~\cite{MoTaNe2008}).
The second reason is as follows. Because $m_M \approx 0.221 \, m_e$ is
below the electron mass, the effect of the virtual particle on the electronic
hydrogen spectrum is no longer parametrically suppressed.  A calculation shows
that because the hypothetical virtual particle now is ``too light,'' the proton
radius inferred from the hydrogen spectroscopy experiments would increase to
$0.939(7) \, {\rm fm}$, because of the concomitant Dirac $\delta$-like vacuum
polarization potential induced by the light millicharged particle.  This is
{\em a priori} not a problem, because the same modification (due to the
millicharged particle) would have to be applied to the proton radius inferred
from scattering experiments~\cite{BeEtAl2010}. However, in that case, since the
discrepancy $\delta E$ is to be explained by the millicharged particle, the
proton radius inferred from the {\em muonic} hydrogen Lamb shift would be equal
to the CODATA median value of $r_p = 0.8768 \, {\rm fm}$, and there would thus
be a $9\,\sigma$ deviation from the modified value $r_p = 0.939(7) \, {\rm fm}$
inferred from electronic hydrogen spectroscopy.  Therefore, the parameter range
$\alpha m_e \ll m_M \ll m_e$ for the millicharged particle also can be excluded.

%
% Unstable Millicharged Particles and Neutral Vector Bosons
%
\subsection{Unstable Neutral Vector Bosons}
\label{sec43}

As evident from Figs.~\ref{abb4} and Fig.~\ref{abb5},
our attempts to explain both muonic QED discrepancies with the simple
form~\eqref{deltarho} of the modified vacuum polarization charge density have
not proven successful. In principle, one may justify a more
complicated {\em ansatz} for the modification, and with enough 
free parameters, it will certainly be possible to 
find a convenient representation that ``explains'' both 
discrepancies and is not in conflict with the electronic
hydrogen Lamb shift and with the electron anomalous magnetic 
moment. This is not our goal.

However, one further, specific form of a modified spectral
density of the vacuum polarization deserves a discussion.
A light, neutral vector boson has been investigated 
as a possible candidate to explain a prevailing
discrepancy of the decay rate of orthopositronium 
(experiment versus theory, see Refs.~\cite{MiEtAl1993,MiEtAl1996,Cz1999}),
which has eventually been resolved~\cite{VaZiGi2003}.
Just like the $\rho(770)$ vector boson, such a hypothetical, virtual
neutral vector boson would induce a small hump in 
the spectral density of vacuum polarization,
corresponding to a resonance in the
photon propagator. The modification
would be restricted to a finite subinterval of the $t$
parameter.  This possibility is not absolutely excluded by other 
searches~\cite{MiEtAl1993,MiEtAl1996}
because the vector boson might be unstable.
We have performed extensive numerical experiments,
for masses (and widths) of the virtual vector boson in the 
primary range $m_e \ll m_V \ll m_\mu$.
One example is 
\begin{equation}
\label{deltarho2}
\delta \rho(t) = \frac{1}{30} \, 
\Theta(t - 20 \, m_e^2) \, \Theta( 24 m_e^2 - t) \,.
\end{equation}
For this choice, 
the muon anomalous magnetic moment discrepancy is reduced to
$2.5\,\sigma$, with the shifted theoretical prediction now lying above the
experimental value. In addition, the muonic hydrogen Lamb shift discrepancy is
reduced to $2.6 \sigma$, with the theoretical value still lying below the
experimental one. The {\em muonic} hydrogen value of the proton radius would
thus shift to a value of $r_p = 0.858 \, {\rm fm}$.  
The electron anomaly $a_e$ is shifted by 
$4.2 \times 10^{-14}$ which is below the current experimental 
uncertainty~\cite{HaFoGa2008}.
However, an evaluation of the
additional vacuum polarization correction due to $\delta \rho$ as given in
Eq.~\eqref{deltarho2} for {\em electronic} hydrogen shows that the proton radius
inferred from hydrogen spectroscopy would also have to be modified, namely to
$r_p = 0.904(7) \, {\rm fm}$, because of the additional Dirac $\delta$
potential induced by the virtual vector boson.  This would leave a
dissatisfactory $6.7 \sigma$ deviation between the modified radii from the two
bound systems. Despite extensive numerical experiments, we
have not found a simple, satisfactory parameter combination that might 
explain both muonic QED discrepancies without significantly distorting the 
proton radius inferred from electronic hydrogen.

%
% Formation--Process Dependent Screening Corrections
%
\section{Formation--Process Dependent Screening Corrections}
\label{straw}

All hypothetical explanations for the discrepancy of theory and experiment in
muonic hydrogen discussed so far in this article
do not seem to lead
to a satisfactory explanation.  In order to understand a physical problem, null
results can also be important, but are not gratifying and leave the effect
unexplained. So, in order to fully understand the problem, we also
analyze the experimental procedures used in Ref.~\cite{PoEtAl2010}.

In the experiment~\cite{PoEtAl2010}, devices have been installed
in the beam line to extract electrons. 
E.g., as revealed in Fig.~2 of Ref.~\cite{PoEtAl2010}, the beam line is 
designed so that muons pass two stacks of thin carbon foils, and 
the electrons released {\em from the foils} are then 
extracted from the beam line via $\vec E \times \vec B$ drift and 
detected in scintillators. This leads to a separation of muons and 
electrons, and only muons continue in the beam line,
to hit the gas target. The molecular hydrogen gas target 
is installed in the beam line {\em behind} the foils, 
i.e., {\em after} the electrons have been extracted from the beam
(but not from the hydrogen molecules in the gas target).

In many other experiments involving the high-precision spectroscopy of muonic
transitions, the electron screening correction has been the limiting factor in
analyzing the experiments~\cite{DuEtAl1974,BeEtAl1986}. Quite exotic 
processes involving muonic atoms have been studied in the literature 
(see, e.g., Ref.~\cite{GrWyBrMu1995}). One of the most striking surprises
is the role of exotic bound states like the well-known~\cite{Ve1967}
molecular state composed of a $pp \mu^-$ ``nucleus'', 
and another proton, bound together by two orbiting electrons.
The mentioned state is known to play a role in muon-catalyzed fusion.
One might thus ask
to which extent the electrons in the ${\rm H}_2$ gas target may influence the
observed lines. Note that the formation process of muonic hydrogen
is complex and it is nontrivial to exclude the contribution of 
resonances from other muonic bound states; for this reason,
resonances of the muonic molecules $pp \mu$ have been
studied~\cite{KiKaHi2004}. The first excited state of the $pp \mu$ molecule
is predicted to have a lifetime of $0.0713 \,{\rm ps}$, close to half that
of a muonic hydrogen atom in the $2P$ state.  As the width of the resonance
observed in the experiment~\cite{PoEtAl2010} is close to the calculated width
for the muonic hydrogen $2P$ state, the authors of~\cite{PoEtAl2010} exclude the 
possibility that the molecular
resonance may contribute to the observed signal.

However, these considerations do not exclude 
contributions from $p \mu^- e^-$ atoms composed of 
a proton, a negative muon and an electron, which might be formed 
in the gas target (these would be heavy analogues of the hydrid ion $H^-$).
Because the muon's orbit is close to the proton,
the proton charge in $p \mu^- e^-$ is shielded from the outer electron.
However, the muon and proton in the $2S$ or $2P$ state form a neutral core 
that interacts with the electron via dipole interactions.
It is known that an ion-atom interaction with a 
functional dependence of the form $-1/r^4$ can form bound 
states (for a recent numerical investigation, see Ref.~\cite{Ga2010}).
Here, the ``ion'' is the electron, whereas the ``atom'' is the 
muon-proton core.
The form of the interaction potential can be derived as follows.

We first slightly generalize the problem and assume that the 
proton is a nucleus with charge number $Z$.
If we have a nucleus with charge number $Z$ and a muon
and an electron bound to it, then
the unperturbed Hamiltonian for the muon reads
\begin{equation}
H_\mu = 
\frac{\vec p_\mu^{\,2}}{2 m_\mu} - \frac{Z\alpha}{r_\mu}  \,,
\end{equation}
and the unperturbed Hamiltonian for the electron is 
\begin{equation}
H_e = \frac{\vec p_e^{\,2}}{2 m_e} - \frac{(Z-1)\alpha}{r_e}  \,.
\end{equation}
because the outer electron merely sees the screened charge of the inner
core, which is the nucleus of charge number $Z$
plus the negative muon of negative charge,
effectively reducing $Z$ by one unit.
For a molecular hydrogen gas target, we can set $Z=1$ (approximately), and 
$H_e$ is approximately equal to the free Hamiltonian but will
be corrected by a $-1/r^4$ interaction, as detailed below.
The total Hamiltonian, including the muon-electron repulsion, then is
\begin{equation}
H = 
\frac{\vec p_\mu^2}{2 m_\mu} - \frac{Z\alpha}{r_\mu} +
\frac{\vec p_e^2}{2 m_e} - \frac{Z\alpha}{r_e} +
\frac{\alpha}{ | \vec r_e - \vec r_\mu |} \,.
\end{equation}
The perturbation is
\begin{equation}
H_I = H - H_e - H_\mu = - \frac{\alpha}{r_e} +
\frac{\alpha}{ | \vec r_e - \vec r_\mu |} \,.
\end{equation}
We expand this expression up to the dipole term
and assume that $r_e > r_\mu$, and obtain the 
interaction Hamiltonian,
\begin{equation}
H_I \approx - \frac{\alpha}{r_e} + \frac{\alpha}{r_e}
+ \alpha \frac{r_\mu}{ r_e^2 } \, \hat x_e \cdot \hat x_\mu \,,
\end{equation}
where $\hat x_e = \vec r_e/r_e$, and
$\hat x_\mu = \vec r_\mu/r_\mu$.
The first two terms cancel, and the third gives the 
dipole interaction.
We start from unperturbed states with the electron
in a state with quantum numbers $|n_e \; \ell_e \; m_e \rangle$ and 
the muon in the $| 2S \rangle_\mu$ state.
A calculation in second-order perturbation theory then 
gives the energy perturbation
\begin{align}
\label{cricket}
\delta E =& \; 
-\frac{\alpha^2}{2} 
\left< n_e \; \ell_e \left| \frac{1}{r_e^4} \right| n_e \; \ell_e \right>_e \;
\left\{ \frac23 \, \sum_{i=1}^3 \sum_{n_\mu} \sum_{m_\mu=-1}^1
\frac{ | \left< 2S \left| r_\mu^i \right| n_\mu \, P \, m_\mu \right>_\mu |^2}%
{E^{(\mu)}_{n_\mu \, P} - E^{(\mu)}_{2S}} \right\} \,.
\end{align}
Here, the superscripts and subscripts identify the particles ($e$ stands for the 
electron, and $\mu$ stands for the muon). The $| 2S \rangle_\mu$ state
undergoes virtual transitions to muonic $|n_\mu P \rangle$ states, and 
the term in curly brackets in Eq.~\eqref{cricket} is recognized 
as the static polarizability of the $2S$ state.
The energetically closest state to the 
muonic $2S$ state is the $| 2P_{1/2} \rangle_\mu$ state, which is energetically 
removed from $| 2S \rangle_\mu$ only by the Lamb shift.
The energy difference is of the order
\begin{equation}
E_{2 P_{1/2}}^{(\mu)} - E_{2S}^{(\mu)} \sim \alpha^3 m_\mu \,.
\end{equation}
The dipole matrix elements for the virtual transitions of the 
muon are of the order of
\begin{equation}
\left< 2S \left| r_\mu^i \right| n_\mu \, P \, m_\mu \right> \sim
\frac{1}{\alpha \, m_\mu} \,.
\end{equation}
We assume that the electron is in a (superposition of) states with quantum numbers 
$| n_e \; \ell_e \; m_e\rangle$ whose dimensions are of the 
order of the (electronic) Bohr radius, and so
\begin{equation} 
\left< n_e \; \ell_e \left| \frac{1}{r_e^4} \right| n_e \; \ell_e \right> 
\sim (\alpha m_e)^4 \,.
\end{equation}
The result of our order-of-magnitude estimate is
\begin{equation}
\delta E \sim -\alpha^2 \, (\alpha m_e)^4 \, \left( \frac{1}{\alpha^5 m_\mu^3} \right) = 
-0.42 \, {\rm meV} 
\end{equation}
for the energy shift of the $2S$ state, if an additional electron is 
present. As this is the lower state of the $2P$--$2S$ Lamb shift
transition, the negative of $\delta E$, i.e., $+ 0.42 {\rm meV}$,
needs to be added to the transition frequency, potentially
explaining the discrepancy. The corresponding effect,
evaluated for the muonic $2P_{1/2}$ state, has the 
opposite sign and therefore shifts the 
transition in the same direction as the effect calculated here.

{\em The main result of the above order-of-magnitude estimate is as 
follows. If, for some reason, an electron is spatially displaced from the 
muonic hydrogen atom by only a few Bohr radii
at the time of the laser-induced Lamb shift transition,
then the static polarizability of the $2S$ muonic state induces
a systematic shift of the muonic transition that has the
right sign and might be large
enough to explain the observed discrepancy.
In order to assess the viability of the $p \mu^- e^-$ atom 
hypothesis, one would have to calculate its spectrum,
its ionization cross sections 
in collisions in collisions with other hydrogen molecules
in the gas target.
Furthermore, it would be necessary to study
its inner Auger rates of $p \mu^- e^-$ as a function of the state of the outer electron,
and its production cross sections in the collisions that take place in the 
molecular hydrogen target used in the experiment~\cite{PoEtAl2010}.
This is beyond the scope of the current article.
The occupation numbers may depend on the formation process.
In order to explain the single, well defined resonance line 
seen in the experiment~\cite{PoEtAl2010}, one would have to assume 
that formation proceeds predominantly into a specific state
of the $p \mu^- e^-$ atom (or into a sufficiently narrow subset
of resonances). These requirements may, in the end, reveal that
the current hypoethesis cannot explain the observed shift
of the resonance line from the predictions of QED theory.
Thus, we do not claim that the calculated effect necessarily 
explains the discrepancy of theory and experiment. 
However, as we were unable to discern viable theoretical
explanations, our general statement is that it may be worthwhile 
to study possible explanations
based on a process-dependent effect. }

Another mechanism by which an electron screening correction could enter the
analysis of the experiment~\cite{PoEtAl2010} might be from neighboring hydrogen
molecules or atoms.  In general, one assumes~\cite{Wi1950} that the initial
capture into highly excited states with principal quantum number 
$n \approx \sqrt{m_\mu/m_e} \approx 14$
takes place via the reaction $\mu^- + {\rm H}_2 \rightarrow (p p \mu^- e^-) +
e^-$, where the muon is captured into a molecular orbit and an electron is
ejected.  Auger rates induced by neighboring hydrogen atoms and molecules
dominate in the deexcitation process for liquid hydrogen~\cite{LeBe1962}, but are
suppressed for a less dense hydrogen gas target.  The deexcitation and muonic
hydrogen formation process has been analyzed both
theoretically~\cite{Wi1950,LeBe1962,BoHeLW1996,ByCzPo1999,JeMa2002} as well as
experimentally~\cite{PoEtAl2006,LuEtAl2007}. Still, there is a
multitude of available reaction channels and conceivable bound states.
It may thus be worthwhile to reexamine the dynamics of the
formation process of muonic hydrogen in the $5 \,{\rm T}$ magnetic field with a
special emphasis on the hypothetical existence of neighboring,
screening electrons that may perturb the
observed frequency in the experiment~\cite{PoEtAl2010}.

%
% Conclusions
%
\section{Conclusions}
\label{conclu}

Recently, two serious discrepancies of theory and experiments
have been observed 
for muonic QED systems: the muon $g$ factor discrepancy of 
$3.4 \, \sigma$ appears to 
persist~\cite{BeEtAl2006muon,BeEtAl2002posmuon,BeEtAl2004negmuon,HaMaNoTe2007},
and for the muonic hydrogen Lamb shift~\cite{PoEtAl2010}, 
an even larger discrepancy is observed which  amounts to $5.0 \, \sigma$.
The proton radius inferred from the 
muonic hydrogen Lamb shift is  $5.0\,\sigma$
smaller than that 
determined by electronic hydrogen spectroscopy~\cite{MoTaNe2008},
or alternatively,
if the proton radius from hydrogen spectroscopy
or from electron scattering is used in order to 
obtain a theoretical prediction for muonic hydrogen,
we observe a $5.0\sigma$ discrepancy of theory and experiment.
The muonic Lamb shift discrepancy is statistically more 
significant and thus, arguably, more urgent to be resolved than the 
muon $g$ factor. Proton structure effects (discussed
in Sec.~\ref{sec3}) and small modifications of the 
vacuum polarization potential (see Sec.~\ref{sec4})
might be discussed as hypothetical explanations for the observed
discrepancy.

As shown in Sec.~\ref{sec3}, an elaborate reevaluation
of the two-photon exchange graph for muonic hydrogen,
including its inelastic (proton polarizability) contribution,
might contribute to an explanation of the muonic Lamb shift 
discrepancy. The inelastic part has been 
evaluated~\cite{Pa1999,Ro1999,MaFa2000} to be 
on the order of
$0.0015(4)\,{\rm meV}$. Unless a reevaluation reveals 
a somewhat surprising enhancement of the contribution by at least an order of 
magnitude, the discrepancy of $\delta E = 0.31 \,{\rm meV}$ will persist.
An alternative explanation
due to a conceivable ``dip'' in the slope of the proton
form factor in the momentum range probed via muonic hydrogen
spectroscopy (see Sec.~\ref{sec31}) could only be excluded 
conclusively by a measurement of the proton 
form factor in scattering experiments probing the momentum
range indicated in Eq.~\eqref{probemuon}.
While certain dip and hump structures in the proton
form factor have been seen in experiments~\cite{BeEtAl2010} and 
in theoretical calculations~\cite{BeHaMe2006}, 
one has to admit that the ``dip'' hypothesis seems somewhat remote.

In Sec.~\ref{sec4}, we investigate the role of a hypothetical millicharged
particle in a numerically small, but important modification of the spectral
density of vacuum polarization. We focus on millicharged particles and 
do not treat hypothetical supersymmetric models. 
The conclusion is as follows: The most
immediate theoretical {\em ansatz} for the modification $\delta \rho(t)$ given
in Eq.~\eqref{deltarho} comes from a low-energy millicharged particle; a model
for a hypothetical virtual natural vector boson resonance is described by
Eq.~\eqref{deltarho2}. Our numerical experiments show that neither a simple
modification of the vacuum polarization function due to a millicharged particle
nor due to an unstable intermediate vector boson can simultaneously explain
both discrepancies observed for the muon anomalous magnetic moment and for the
muonic hydrogen Lamb shift without significantly distorting the proton radius
inferred from electronic hydrogen.  Millicharged particles and virtual vector
bosons could explain the muon anomalous magnetic discrepancy while having a
negligible effect on the muonic Lamb shift, but not vice versa.  If these
particles are unstable, then other parametric bounds (e.g., those following
from the experimental investigation reported in Ref.~\cite{PrEtAl1998}) may not
be applicable to an analysis of their {\em virtual} contributions within vacuum
polarization loops.

Our investigations severely restrict the parameter space available for a
modification of the vacuum polarization spectral function due to either
millicharged particles or unstable virtual vector bosons in the low-energy
domain. Being statistically (barely) significant, the muon anomalous magnetic
moment discrepancy is numerically so small that it restricts possible
modifications of the vacuum polarization spectral function due to hypothetical
millicharged particles to such small coupling strengths that a simultaneous
explanation of the comparatively large observed discrepancy in muonic hydrogen
becomes impossible.  In a small parameter range, a virtual vector boson
resonance modification somewhat reduces the statistical significance of the
combined discrepancies in muonic hydrogen and for the muon anomalous magnetic
moment, but at the cost of inducing a prohibitively large modification of the 
(electronic) hydrogen Lamb shift.

Hitherto undetected virtual particles
from extensions of the standard model might still
explain the two 
discrepancies $\delta E$ and $\delta a_\mu$,
but a more complex structure of the modification of the
vacuum polarization spectral function would have to be generated, 
with---possibly---both attractive as well
as repulsive modifications.
Repulsive modifications of the 
vacuum polarization potential have been discussed
in the literature~\cite{Ad1974prd1,Ad1974prd2}. A more complex modification of the vacuum
polarization has more free parameters.
Therefore, a conceivable determination of these parameters 
becomes possible, if at all, only when more experimental
data on muonic systems (e.g., the muonic helium ion) become available.
Our discussion in Secs.~\ref{sec42} and~\ref{sec43} does not 
exhaust all possible theoretical explanations from extensions 
of the standard model but covers particular models that 
``suggest themselves.''
A conceivable explanation from an extension of the standard model
is constrained by the comparatively large discrepancy in the 
muonic hydrogen Lamb shift, the comparatively small
discrepancy in the muon $g$ factor, and must respect the 
excellent agreement of theory and experiment for the 
electron $g$ factor and the hydrogen Lamb shift.

Having discussed the most attractive and far-reaching theoretical
consequences of the recently observed discrepancy~\cite{PoEtAl2010}
and having found negative results, we then proceed to indicate 
a candidate effect for a process-dependent correction to the energy 
levels that may be relevant to the experiment~\cite{PoEtAl2010}
(see Sec.~\ref{straw}). We believe that a study of exotic bound states
and a reexamination of the dynamics of the formation process
indicated in Sec.~\ref{straw} might be interesting in its own right 
even if eventually, more detailed calculations might show that these effects
do not offer the explanation for the observed discrepancy in 
muonic hydrogen. 

High-precision QED experiments are excellent probes of conceivable low-energy
modifications of the standard model.  Our considerations highlight the need for
further experimental evidence regarding muonic bound systems, before conclusive
statements are possible.
As discussed in Sec.~\ref{review}, discrepancies of theory and 
experiment in muonic systems have been encountered in the past 
and have eventually been resolved. 

Even if theoretical considerations and careful considerations of 
additional systematic effects concerning the experiment~\cite{PoEtAl2010}
fail to resolve the proton radius discrepancy, then there is a third way 
to resolve it, based on spectroscopic methods. It works as follows.
In general, the fundamental constants derived from atomic spectroscopy
are highly intertwined (not only the nuclear radii). One
example is the Rydberg constant. If the new proton radius
from Ref.~\cite{PoEtAl2010} is inserted into the evaluation of
electronic hydrogen spectra, then a very precise Rydberg
constant is obtained, which is given as~\cite{PoEtAl2010}
\begin{equation}
\label{cR1}
c \, R_\infty = 3\,289\,841\,960\,251(5) \, {\rm kHz} \qquad 
[1.5 \times 10^{-12}]\,,
\end{equation}
where we indicate the relative uncertainty in square brackets.
In an unpublished PhD thesis~\cite{dV2002} completed at MIT (Cambridge)
in 2002,
based on Rydberg transitions which are manifestly independent of the
proton charge radius, the value
\begin{equation}
\label{cR2}
c \, R_\infty = 3\,289\,841\,960\,306(69) \, {\rm kHz} \qquad 
[2.1 \times 10^{-11}]\,,
\end{equation}
is obtained [the difference to the value in Eq.~\eqref{cR1} is  $0.8 \sigma$].
The CODATA value is
\begin{equation}
\label{cR3}
c \, R_\infty = 3\,289\,841\,960\,361(22) \, {\rm kHz} \qquad 
[6.6 \times 10^{-12}]\,,
\end{equation}
and the difference to~\eqref{cR2} also is $0.8\sigma$, but ``to the
other side.'' The Rydberg state value~\eqref{cR2} lies in between the values
given in Eqs.~\eqref{cR1} and~the CODATA value~\eqref{cR3}.
Quite surprisingly, an improved measurement of the Rydberg
constant based on ionic Rydberg states, though in itself
independent of the proton radius,  could thus settle the
proton radius question connected with the muonic hydrogen measurement.
Such a measurement is currently being pursued at the National Institute
of Standards and Technology~\cite{JeMoTaWu2008}.

%
% Acknowledgments
%
\section*{Acknowledgments}

The author gratefully acknowledges informative and insightful
discussions with K.~Pachucki and P.~J.~Mohr, and warm hospitality at the National
Institute of Standards and Technology in August 2010,
where an important part of this work has been performed.
Many helpful conversations with J.~Sapirstein,
S.~J.~Brodsky and U.~P.~Jentschura are also gratefully acknowledged.
This research has been supported by a NIST Precision Measurement Grant and
by the National Science Foundation.

\rhead{\sc References}

%{\em Literature references have been updated at the time of submission
%of the current paper to a scientific journal.}

\end{document}